\documentclass[sigconf,screen]{acmart}

\newcounter{BalanceAtReference}
\newcounter{ReferenceIndexForBalancing}

\makeatletter

\global\@ACM@balancefalse

\def\@balancelastpageonce{%
  \ifnum\value{ReferenceIndexForBalancing}=\value{BalanceAtReference}
    \newpage
  \else
    \relax
  \fi
  \stepcounter{ReferenceIndexForBalancing}
}
\pretocmd{\bibitem}{\@balancelastpageonce}
  {} %
  {\@latex@error{Patching \bibitem failed}{\@ehd}}

\makeatother

\setcopyright{rightsretained}
\acmPrice{}
\acmDOI{10.1145/3540250.3549082}
\acmYear{2022}
\copyrightyear{2022}
\acmSubmissionID{fse22main-p38-p}
\acmISBN{978-1-4503-9413-0/22/11}
\acmConference[ESEC/FSE '22]{Proceedings of the 30th ACM Joint European Software Engineering Conference and Symposium on the Foundations of Software Engineering}{November 14--18, 2022}{Singapore, Singapore}
\acmBooktitle{Proceedings of the 30th ACM Joint European Software Engineering Conference and Symposium on the Foundations of Software Engineering (ESEC/FSE '22), November 14--18, 2022, Singapore, Singapore}

\AtBeginDocument{%
  \providecommand\BibTeX{{%
    \normalfont B\kern-0.5em{\scshape i\kern-0.25em b}\kern-0.8em\TeX}}}

\input{utils.tex}
\usepackage{xcolor} %
\definecolor{darkgreen}{rgb}{0.05,0.5,0.05}

\newcommand{\gh}{GitHub\xspace}
\newcommand{\sic}{\emph{[sic]}\xspace}

\newcommand{\category}[1]{\textbf{#1}}
\newcommand{\subcategory}[1]{\emph{\faCaretRight{}#1}}

\newcommand{\RQA}{What are relevant skills for OSS?\xspace}
\newcommand{\RQB}{How do OSS contributors grow and improve skills?\xspace}
\newcommand{\RQC}{How do OSS contributors share and display skills?\xspace}

\RequirePackage{color}
\definecolor{BLUE}{rgb}{0,0,1}
\definecolor{BLACK}{rgb}{0,0,0}
\definecolor{HIGHLIGHTCOLOR}{rgb}{0,0,0}

\newcommand{\ADDED}[1]{{\protect\color{HIGHLIGHTCOLOR}#1}}

\usepackage{multirow}
\usepackage{multicol}

\begin{document}

\title{Understanding Skills for OSS Communities on GitHub}

\author{Jenny T. Liang}
\affiliation{%
  \institution{University of Washington}
  \city{Seattle}
  \state{Washington}
  \country{USA}
}
\email{jliang9@cs.washington.edu}

\author{Thomas Zimmermann}
\affiliation{%
  \institution{Microsoft Research}
  \city{Redmond}
  \state{Washington}
  \country{USA}}
\email{tzimmer@microsoft.com}

\author{Denae Ford}
\affiliation{%
  \institution{Microsoft Research}
  \city{Redmond}
  \state{Washington}
  \country{USA}}
\email{denae@microsoft.com}

\renewcommand{\shortauthors}{Liang, et al.}

\begin{abstract}
The development of open source software (OSS) is a broad field which requires diverse skill sets. For example, maintainers help lead the project and promote its longevity, technical writers assist with documentation, bug reporters identify defects in software, and developers program the software.
However, it is unknown which skills are used in OSS development as well as OSS contributors' general attitudes towards skills in OSS.
In this paper, we address this gap by administering a survey to a diverse set of 455 OSS contributors.
Guided by these responses as well as prior literature on software development expertise and social factors of OSS, we develop a model of skills in OSS that considers the many contexts OSS contributors work in. This model has 45 skills in the following 9 categories: technical skills, working styles, problem solving, contribution types, project-specific skills, interpersonal skills, external relations, management, and characteristics.
Through a mix of qualitative and quantitative analyses, we find that OSS contributors are actively motivated to improve skills and perceive many benefits in sharing their skills with others. We then use this analysis to derive a set of design implications and best practices for those who incorporate skills into OSS tools and platforms, such as \gh. 
\end{abstract}

\ccsdesc[500]{Human-centered computing~Open source software}

\keywords{open source software, skills, empirical study, survey}

\maketitle

\section{Introduction}

The development of open source software (OSS) is a complex undertaking, which requires a diverse set of people and skill sets. Successful OSS projects are not completed alone: maintainers drive the project to be sustained long-term, keeping in mind their community and project vision; technical writers generate clear, well-written documentation to teach others to use the OSS; bug reporters identify defects which improve the product; and software developers help program software, create new features, and fix bugs. 

At the crux of these roles are \emph{skills}. Skills are critical to software development, including the development of OSS. While OSS development and software development are similar, OSS contributors also work in contexts that are unique to OSS (e.g., consistently collaborating in a distributed form). Previous work has investigated the experience, expertise, and skills of OSS contributors in a variety of contexts, such as maintainers~\cite{dias2021makes}, integrators~\cite{gousios2015work}, and mentors~\cite{balali2018newcomers, steinmacher2021being}. While work such as these provide some idea of skills that are necessary for OSS, it does not fully consider the broad set of circumstances that OSS contributors work in, such as contributors who provide essential non-code contributions~\cite{allcontributors2022}. Prior work also does not provide a set of skills relevant to OSS and furthermore, how contributors would like to share and improve on them. 

By better understanding skills related to OSS, we can inform the design of tools or platforms that incorporate skills and elicit best practices to leverage them as much as possible. In this paper, we investigate the skills which are relevant to OSS and contributors' relationship with skills.
We define three research questions:

\begin{description}
\setlength{\itemsep}{0pt}
\setlength{\parskip}{2pt}
    \item[\textbf{RQ1}] \RQA
    \item[\textbf{RQ2}] \RQB
    \item[\textbf{RQ3}] \RQC
\end{description}

To answer the research questions, we deployed a survey to elicit OSS contributors' insights and to understand their preferences related to skills. Our survey received 455 responses with perspectives from OSS contributors.
We derive a set of relevant skills from the survey responses and by leveraging the rich set of literature surrounding social factors of OSS and software development expertise. From the survey data, we further investigate contributors' perspectives on skills in OSS, namely how they would like to improve skills and how they would like to share their skills.

In our study, we find that contributors themselves defined skills in OSS very broadly, which is amplified by the diverse set of roles in OSS. Thus, in this paper, we use a general notion of skills, also following prior work~\cite{vadlamani2020studying}. \textbf{We define skills as anything that an individual can actively work to improve on.}
Our definition encompasses formal skills as well as personal attributes of contributors. %
The broader definition of skill is exemplified by this response: 
\vspace{-6pt}
\quo{Skills are anything that helps improve the project. This can include coding, writing docs, researching bugs, triaging incoming bug reports and requests, and doing releases. Larger projects may also require skills that we see in commercial dev environments, such as project management, product management, and planning release schedules. And there is also a need for people skills like dealing with conflict between project members, enforcing codes of conduct, and other skills along those lines.}{R_Ath7Jwsh93gQzh7}
\vspace{6pt}

We find that OSS skills include a healthy mixture of hard skills (e.g., software engineering and other technical expertise) and soft skills (e.g., interpersonal and management skills). Furthermore, we find that OSS contributors are actively motivated to improve skills and perceive many benefits in sharing their skills with others. However, sharing skills may also have potential negative effects, which deserves careful consideration.

The contributions of this paper are as follows:
\begin{itemize}
    \item A model of skills in OSS informed by literature and practitioner experiences \ADDED{from 455 survey responses} %
    (Section~\ref{sec:skills}).
    \item Survey findings on how contributors would like to improve and share skills (Sections~\ref{sec:improving} and~\ref{sec:showing}).
    \item Best practices and design implications on incorporating skills into OSS platforms, such as \gh (Section~\ref{sec:Discussion}).
\end{itemize}

\section{Background}
Software engineering researchers have studied the skills and expertise that software engineers rely on. Previous work has found that effective soft skills (e.g., communication, critical thinking)~\cite{ahmed2012evaluating, florea2018software, montandon2021skills} and hard skills (e.g., fault localization, software design)~\cite{xia2019practitioners} are crucial to software engineering. Baltes et al. developed a theory of software engineering expertise with a focus on programming expertise being formed~\cite{baltes2018towards}. Li et al. uncovered 56 attributes of great software engineers, focusing on their soft skills, such as personal and decision making characteristics~\cite{li2015makes}. \ADDED{In a later study, Li et al. found that distinguishing traits of great software engineers included paying attention to coding details, handling complexity, and continuously improving as well as being open-minded and honest~\cite{li2020distinguishes}.} However, these models cannot be fully applied to OSS development. OSS development relies on roles aside from software engineering, such as community managers, technical writers~\cite{trinkenreich2020hidden}, and people with external domain expertise~\cite{huang2021leaving}. Thus, it has unique contexts (e.g., wrangling contributors~\cite{dias2021makes, gousios2015work}, identifying funding~\cite{dias2021makes, trinkenreich2020hidden}, consistently collaborating in a distributed, remote form~\cite{gousios2015work, gousios2016work}) which prior research in this area does not completely address. 

\ADDED{Practitioners have previously discussed the various skills required in OSS communities. This includes social skills, such as communication, collaboration, and advocacy, as well as technical skills, such as tools and technologies, programming, writing documentation, and bug triaging~\cite{fogel2005producing,brasseur2018forge,bacon2012art}. Learning about OSS and related skills have also occurred in academic settings. For instance, universities have developed graduate degree programs and courses on free open-source software for students to become experts in the field~\cite{megias2005international,montes2014considerations}. Students have also been invited to participate in OSS projects as part of class assignments to teach collaborative software engineering skills~\cite{sowe2007involving,kilamo2012teaching} and knowledge in OSS work practices and development processes~\cite{lundell2007learning}}.

\ADDED{Additionally, researchers have investigated contributor motivation. Contributors participate in OSS projects for self-marketing and to fulfill personal software needs~\cite{alexander2002working} as well as having fun and giving back to others.~\cite{bitzer2007intrinsic}. Huang et. al found that contributors involved in OSS projects for social good were motivated to join to impact society and learn new skills~\cite{huang2021leaving}. Other research has also studied career trajectories of contributors. Trinkenreich et al. found that contributors often start their OSS career by ``lurking" in the community, such as working with the code or joining events. Then, as involvement increased, they would take on coding or non-coding roles. Contributors' careers were non-linear; they could switch roles throughout their careers.~\cite{trinkenreich2020hidden}}

Meanwhile, previous work in OSS has uncovered aspects of being a successful OSS contributor across a variety of specific contexts. However, this body of work does not provide a holistic model of OSS contributor skills or attributes across the OSS ecosystem. For example, previous work has independently studied many roles in OSS, such as attributes of maintainers~\cite{dias2021makes}, barriers of newcomers~\cite{steinmacher2015social, steinmacher2015systematic} and mentors~\cite{balali2018newcomers}, the practices of integrators~\cite{gousios2015work}, and experiences of quasi-contributors (i.e., contributors who did not have their contributions accepted)~\cite{steinmacher2018almost}. Other work has examined how contributions~\cite{tsay2014influence} and other contributors~\cite{marlow2013activity, marlow2013impression} are evaluated by OSS contributors on \gh.

Related to skills in OSS is Trinkenreich et al.'s work on definitions of success in OSS~\cite{trinkenreich2021pot}. Some factors of success includes cooperation, advancement by level and experience, self-development, and achieving personal satisfaction. Our work complements this by enumerating what skills are necessary for contributors to achieve some forms of success based on their model.

\section{Methodology}
\label{sec:methodology}

To answer the three research questions, we conducted a literature review (Section~\ref{sec:LiteratureReview}) and deployed a survey among contributors of open source (Section~\ref{sec:SurveyMethod}). To analyze the data, we used a combination of qualitative coding and statistical analysis (Section~\ref{sec:analysis}). Materials to help with replication of this work are available as supplemental materials~\cite{supplemental-materials}.

\subsection{Literature Review}
\label{sec:LiteratureReview}
To compile a list of papers related to skills in OSS, we searched the ACM Guide to Computing Literature using the following criteria: 
\begin{itemize}
    \item The \emph{content type} was ``Research Article''.
    \item The \emph{title} contained one or more of the following terms: ``open source'', ``oss'', ``software development'', ``developer'', ``developers'', ``software engineer'', ``software engineers''.
    \item The \emph{abstract} contained one or more of the following terms: ``skill'', ``skills'', ``expertise'', ``knowledge'', ``characteristic'', ``attribute''.
    \item The \emph{publication venue} was in a top publkication venue in software engineering (ICSE, ESEC/FSE, FSE, TSE, TOSEM), human-computer interaction (CHI, CSCW), or computer science education (SIGCSE). We included the CHI and CSCW conferences to get an outside perspective of skill and because they regularly publish research related to software engineering. We included the SIGCSE conference because of its focus on computing education, which is related to skills.
    \item The \emph{publication year} was between 2011 and 2021. We selected recent papers to focus on modern OSS tools, technologies, or processes. We selected 2011 as the cutoff year because the first research papers on GitHub and Stack Overflow were published in 2012~\cite{bacchelli2012harnessing,gousios2012ghtorrent,dabbish2012social}.  
\end{itemize}

The initial query yielded 228 results in the ACM Guide.
We excluded papers with fewer than 15 citations unless they had been published recently (2019, 2020, 2021).
The first author then checked each of the papers to see if it included an explicit list of skills, attributes, or characteristics of software developers or OSS contributors.
The result was a set of ten papers
\cite{baltes2018towards, coelho2018we,dias2021makes,groeneveld2020non, hoda2012self, kalliamvakou2017makes, li2015makes, marlow2013activity, mendez2018open, trinkenreich2020hidden}
that discussed skills and expertise in software engineering, often through the lens of social factors of OSS.
The papers covered a diverse set of skills and contribution types. The paper selection was not meant to be exhaustive as we planned to supplement this resource with a survey among OSS contributors.

\begin{figure}
    \begin{tcolorbox}[left=-14pt,right=2pt,top=2pt,bottom=2pt]
    {\sc \hspace{16pt}Survey Questions}
    \begin{itemize}
        \item[{\tiny\faStar}] \questiontext{Q10}
        \item[{\tiny\faStar}] \questiontext{Q11}
        \item[{\tiny\faStar}] \questiontext{Q20}
        \item \questiontext{Q12}
        \item \questiontext{Q14}
        \item \questiontext{Q24}
        \item \questiontext{Q32}
        \item[{\tiny\faStar}] \questiontext{Q33}
        \item[{\tiny\faStar}] \questiontext{Q51}
    \end{itemize}
    \end{tcolorbox}
    \vspace{-0.5\baselineskip}
    \caption{A subset of the survey questions. Open-ended questions are indicated with a star ({\tiny\faStar}). The complete survey instrument is in the supplemental materials~\cite{supplemental-materials}.}
    \label{fig:survey}
\end{figure}

\subsection{Survey}
\label{sec:SurveyMethod}
To understand contributors' perspectives on skills in OSS, we distributed a Qualtrics survey to OSS contributors on \gh.

\paragraph{Design}
We designed a survey to obtain examples of OSS skills and investigate the role of skills in OSS.
Survey topics included: how users currently display skills relevant to OSS, identifying relevant skills in OSS, improving skills in OSS, and evaluating new proposed methods of displaying skills on \gh. 
A subset of the survey questions is displayed in Figure~\ref{fig:survey}; the full survey instrument is available in our supplemental materials~\cite{supplemental-materials}. 

Following previous work~\cite{huang2021leaving}, we used the FDA's descriptions of race and ethnicity~\cite{fda2016collection} and the HCI Guidelines for Gender Equity and Inclusivity~\cite{scheuerman2020hci} to collect demographic information. 
We allowed participants to select multiple responses for the demographics related to ethnicity, gender, and role.
While developing the survey, an external researcher reviewed the survey and provided feedback on the wording of the questions.

\paragraph{Sampling Strategy}
First, we identified \gh projects to sample potential contributors from two lists of OSS projects previously used by Huang et al. to study OSS communities~\cite{huang2021leaving}. The list \emph{Projects-OSS4SG} contained 437 \gh projects identified as OSS for social good (OSS4SG), while the other list \emph{Projects-OSS} contained 642 \gh projects that were not included in the first list.

Next, we defined populations based on their contribution types to each project: 1) authoring commits, 2) opening issues and pull requests, and 3) commenting on others’ issues and pull requests. We chose to sample using different contribution types to capture the diversity of roles that contributors take in OSS projects. We sampled and distributed the survey in two phases.

In \emph{Phase 1}, we focused on commits as a contribution type. We identified the \emph{top 30 contributors by commit count for each project} in \emph{Project-OSS4SG} and \emph{Project-OSS}, which yielded population sizes of 5,905 and 6,256 contributors respectively. 1,856 users from OSS4SG and 2,692 users from OSS had contact information publicly available. After removing duplicates, this resulted in 3,072 users who we invited to take the survey.

In \emph{Phase 2}, we focused on issues and pull requests as contribution types.  We identified all contributors who submitted an issue or opened a pull request in \emph{Project-OSS4SG} (25,011 users) and \emph{Project-OSS} (72,501 users). We also identified all \gh users who commented on another contributor's issue or pull request in \emph{Project-OSS4SG} (5,583 users) and \emph{Project-OSS} (5,065 users). 
From these four lists, we selected users who included contact information in their public \gh profile. We removed duplicate users across the four sets and users who had already been distributed a survey from Phase 1. This resulted in 28,847 users who participated in \gh issues and pull requests. Due to the large size, we further reduced this set to the \emph{users who had opened issues, pull requests, or comments within the past twelve months}. This resulted in 6,023 \gh users who we invited to take the survey.

\paragraph{Participants}
The survey was sent to a total of 9,095 OSS contributors and received 455 responses, resulting in a response rate of 5\%, which is comparable to other surveys in software engineering, which range from 6\% to 36\%~\cite{smith13}. The slightly lower response rate is due to the length of the survey, which was advertised as 20 minutes. Most participants completed the survey within 20 minutes, with a median completion time of 16 minutes. After completing the survey, participants could join a sweepstakes to win one of four \$100 electronic gift cards.

Our participants represented various \emph{geographic locations}, including Africa ($n$ = 23), the Americas ($n$ = 159), Asia ($n$ = 115), Europe ($n$ = 152), and Oceania ($n$ = 6). Participants also represented a wide range of \emph{ethnicities}, including White ($n$ = 229), Asian ($n$ = 123), Black or African ($n$ = 14), Hispanic or Latino ($n$ = 43), Middle Eastern ($n$ = 22), and Native American or Other Pacific Islander ($n$ = 1). Multiple \emph{genders} were also represented, such as woman ($n$ = 26), man ($n$ = 398), and non-binary or gender diverse ($n$ = 9). Participants also were involved in diverse set of \emph{roles} in OSS, such as documentation writer ($n$ = 168), bug submitter ($n$ = 211), graphic artist ($n$ = 5), translator ($n$ = 47), coder ($n$ = 371), maintainer ($n$ = 226), porter ($n$ = 13), tool builder ($n$ = 63), and GUI designer ($n$ = 25).

\subsection{Analysis}
\label{sec:analysis}

To compile the \textbf{list of skills in open source} (RQ1), we included a question in the survey which asked participants to provide examples of skills in OSS, which 368 participants answered. Additionally, from Section~\ref{sec:LiteratureReview}, we compiled a resource which included 193 instances of skills in OSS or software engineering from literature. This yielded a combined dataset of 561 skills-related items to analyze. To remove ordering effects, the dataset was randomly shuffled. 

To analyze the data, all authors followed best practices in qualitative coding and initially coded the same set of statements~\cite{saldana2009coding}. During the first round of qualitative coding, they open-coded the first 100 items from the shuffled dataset. In each item, the authors independently identified relevant OSS skills. Then, they convened to discuss the resulting set of codes and their relationships and to clarify the scope of each code. From this discussion, a shared codebook was generated that included the authors' identified skills. 

Next, the authors independently performed a second round of open coding on the next 100 items using the shared codebook. The authors convened again to discuss the scope of the codes and add new codes by unanimous vote. In the second round, the authors agreed on $78.2$\% of the codes; a moderate agreement which is also comparable to other empirical software engineering studies~\cite{mchugh2012interrater,johnson2016crosstool}. 
Next, the rest of the items were divided into sections and each author coded one of the sections independently. Afterwards, the authors convened for one last discussion on the codes' scopes and added new codes based on unanimous vote, which resulted in the final model in Section~\ref{sec:skills}.

\smallskip

To provide insight on \textbf{improving skills} (RQ2) and \textbf{displaying skills} (RQ3), we analyzed a combination of open-ended and closed-ended questions. For the open-ended questions, each question was assigned a single author. The assigned author then reviewed at least 100 responses and identified common themes across the responses that answered the prompt, each of which corresponded to a code. Finally, each response was labeled with one or more codes. The authors then met to discuss the codes and the emerging themes. %
For the closed-ended questions, we used standard statistical analysis techniques. We report percentages of how frequently an item was selected or how frequently participants agreed or strongly agreed with a statement. This follows the best practices shared by Kitchenham and Pfleeger on how to analyze survey data~\cite{survey-guidelines}.

\begin{figure}
    \centering
    \includegraphics[width=\linewidth]{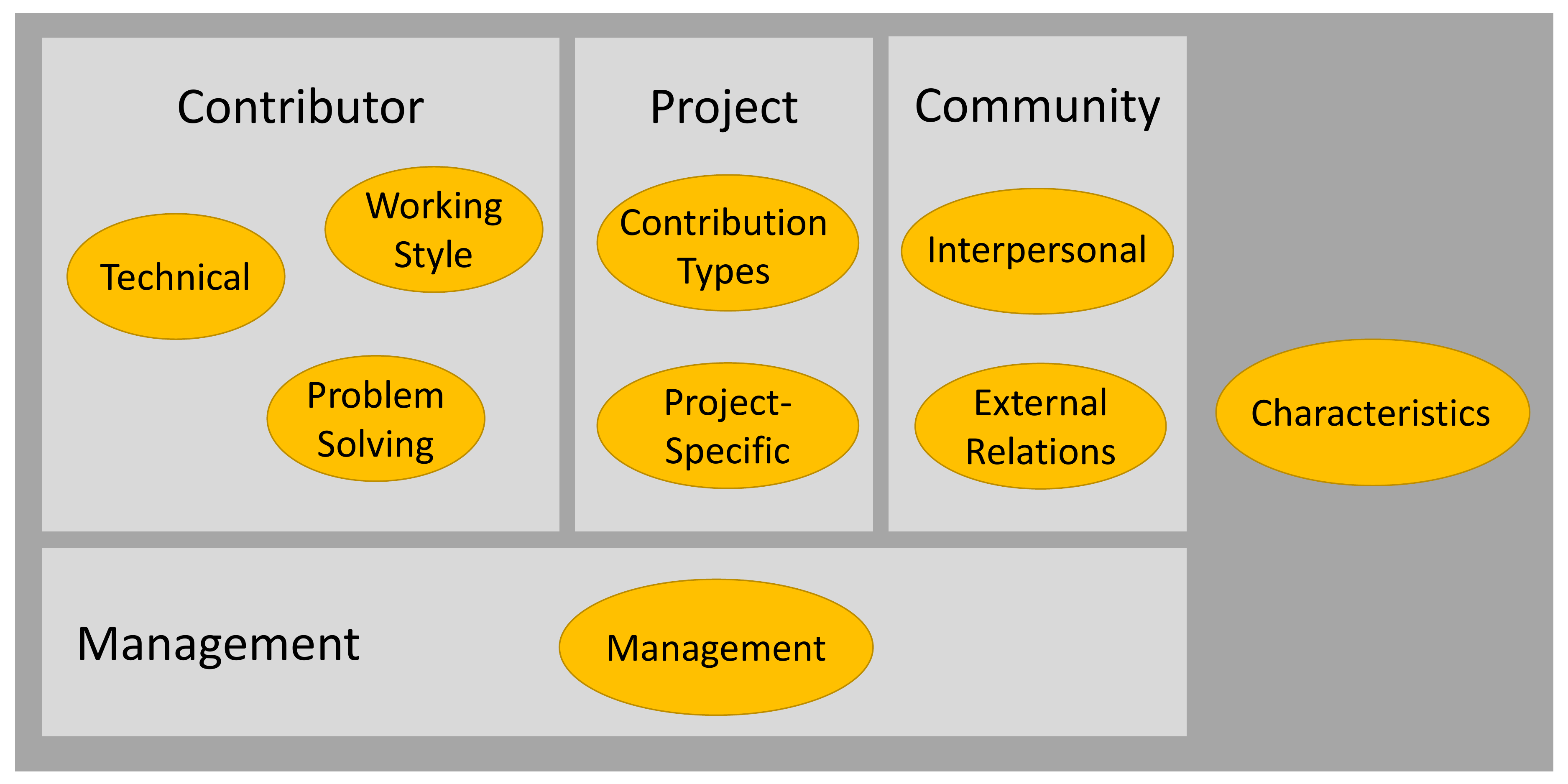}

    \vspace{-0.5\baselineskip}

    \caption{An overview of the skills in our OSS model. Characteristics underlie all skills, while the remaining skill categories directly relate to the contributor, project, or community. Management-related skills cross-cut these functions.}
    \label{fig:ModelOverview}
\end{figure}

\section{Skills in OSS (RQ1)}
\label{sec:skills}

In this section, we present the findings for the research question: \textbf{``\RQA''} 

Through reviewing relevant literature and qualitative coding, we synthesized an OSS skills model with 45 skills across the following 9 categories: technical skills, working styles, problem solving, contribution types, project-specific skills, interpersonal skills, external relations, management, and characteristics (see Figure~\ref{fig:ModelOverview}). Overall,
we find that characteristics of the contributor underlie all other skills and allows an individual to excel in other skills. The remaining skill categories are directly related to either the OSS contributor, project, or community, where management-related skills cross-cut across all three of these functions.

In analyzing survey responses, we find that participants described skills in OSS broadly, and included formal skills as well as personal attributes. Thus, our model includes \emph{any instance of a skill or attribute listed by participants that could be improved upon by an individual}. Below, we discuss this model and its categories in further detail. Our full model is summarized in Table~\ref{tab:SkillsTable}.

\definecolor{cellorange}{rgb}{ 1,  .949,  .8}
\definecolor{cellgreen}{rgb}{ .776,  .878,  .706}
\definecolor{cellblue}{rgb}{ .608,  .761,  .902}
\definecolor{celllightgrey}{rgb}{ .921,  .921,  .921}
\definecolor{celldarkgrey}{rgb}{ .664,  .664,  .664}

\definecolor{shadecolor}{rgb}{.92,  .92, .92} 

\newcommand{\STAB}[1]{\begin{tabular}{@{}c@{}}#1\end{tabular}}

\newcommand{\themeheader}[2]{\textbf{{#2}}}
\newcommand{\skillheader}[1]{\emph{#1}}

\begin{table*}
  \centering
  \caption{A list of skills based on the literature review and analysis of survey responses. \faMale\xspace = 1-10 survey mentions, \faMale\space\faMale\xspace = 11-20 survey mentions, \faMale\space\faMale\space\faMale\xspace = 21-50 survey mentions, \faMale\space\faMale\space\faMale\space\faMale\xspace = 51-100 survey mentions, and \faMale\space\faMale\space\faMale\space\faMale\space\faMale\xspace = 100+ survey mentions.
}

 \resizebox{\linewidth}{!}{%
    \begin{tabular}{p{15cm}p{4.35cm}p{2.25cm}}
    \toprule
 \textbf{\Large Skill \& Description} & \multicolumn{2}{c}{\Large \textbf{Sources}} \\
 \cmidrule{2-3}
 & \hfil Literature & \hfil Survey\\
 \hline

\multicolumn{3}{l}{\cellcolor[rgb]{ .921,  .921, .921} \themeheader{T9}{Technical Skills}}\\
\hline

\skillheader{Programming.} Has skills in basic programming (e.g., testing, code comprehension, debugging) & \hfil \cite{li2015makes}, \cite{mendez2018open}, \cite{trinkenreich2020hidden} & \hfil \faMale \space \faMale \space \faMale \space \faMale \space \faMale  \\
\hline

\skillheader{Software engineering.} Has skills in good software engineering practices (e.g., software design, software architecture, requirements, refactoring, writing maintainable code) & \hfil \cite{baltes2018towards}, \cite{groeneveld2020non}, \cite{li2015makes} & \hfil \faMale \space \faMale \space \faMale \space \faMale \\
\hline

\skillheader{Technologies.} Is familiar with specific programming languages, tools, technologies, or frameworks & \hfil \cite{baltes2018towards}, \cite{kalliamvakou2017makes}, \cite{li2015makes}, \cite{mendez2018open} & \hfil \faMale \space \faMale \space \faMale \\
\hline

\skillheader{DevOps.} Has knowledge in DevOps to deliver software quickly and efficiently & & \hfil \faMale \\
\hline

\skillheader{Domain.} Has domain-specific knowledge or expertise &  \hfil \cite{dias2021makes}, \cite{li2015makes}, \cite{mendez2018open} & \hfil \faMale \space \faMale \\
\hline

\skillheader{Version control systems.} Has basic version control knowledge (e.g., resolving merge conflicts) &  \hfil \cite{mendez2018open} & \hfil \faMale \space \faMale \space \faMale \\[8pt]
\hline

\multicolumn{3}{l}{\cellcolor[rgb]{ .921,  .921,  .921}\themeheader{T9}{Working Styles}}\\
\hline

\skillheader{Excellence.} Strives to achieve excellence or produce high-quality work by paying attention to smaller details; takes pride in one's work &  \hfil \cite{baltes2018towards}, \cite{dias2021makes}, \cite{groeneveld2020non}, \cite{li2015makes}, \cite{marlow2013activity}, \cite{trinkenreich2020hidden} & \hfil  \faMale \space \faMale \\
\hline

\skillheader{Available.} Has time available and is present to answer questions or work with others & \hfil \cite{baltes2018towards}, \cite{coelho2018we}, \cite{dias2021makes}, \cite{kalliamvakou2017makes}, \cite{li2015makes}, \cite{mendez2018open} & \hfil \faMale \\
\hline

\skillheader{Organized.} Keeps things in order; works in a systematic way & & \hfil \faMale \\[8pt]
\hline

\multicolumn{3}{l}{\cellcolor[rgb]{ .921,  .921, .921} \themeheader{T9}{Problem Solving}}\\
\hline

\skillheader{Creative.} Ideates creative solutions to solve a problem & \hfil \cite{groeneveld2020non}, \cite{li2015makes} & \hfil \faMale \\
\hline

\skillheader{Initiative.} Takes initiative to solve problems; executes on the steps to address problems &  \hfil \cite{kalliamvakou2017makes}, \cite{li2015makes}, \cite{marlow2013activity}, \cite{trinkenreich2020hidden} &\hfil \faMale \space \faMale  \\
\hline

\skillheader{Analytical.} Is able to analyze, apply logical reasoning to, or critically think through problems and solutions & \hfil \cite{groeneveld2020non}, \cite{kalliamvakou2017makes}, \cite{li2015makes} & \hfil \faMale \space \faMale \\[8pt]
\hline

\multicolumn{3}{l}{\cellcolor[rgb]{ .921,  .921, .921} \themeheader{T9}{Contribution Types}}\\
\hline

\skillheader{Bug triaging.} Determines the severity of a bug & & \hfil \faMale \space \faMale \\
\hline

\skillheader{Bug reporting.} Finds and reports defects in software & & \hfil \faMale \space \faMale \\
\hline

\skillheader{Code review.} Reads through and provides feedback on code contributions & & \hfil \faMale \space \faMale \\
\hline

\skillheader{Documentation.} Writes content to convey technical information to broad audiences (e.g., documentation) & \hfil \cite{dias2021makes}, \cite{trinkenreich2020hidden} & \hfil \faMale \space \faMale \space \faMale \space \faMale  \\
\hline

\skillheader{Visual design.} Creates visuals for the OSS project (e.g., graphics, UI/UX) & & \hfil \faMale \space \faMale \\
\hline

\skillheader{Translation.} Translates content to another language (e.g., localization) & & \hfil\faMale \\[8pt]
\hline

\multicolumn{3}{l}{\cellcolor[rgb]{ .921,  .921, .921} \themeheader{T9}{Project-Specific Skills}}\\
\hline

\skillheader{Purpose.} Understands and believes in the purpose and societal impacts of the project or OSS development at large & \hfil \cite{dias2021makes}, \cite{kalliamvakou2017makes}, \cite{li2015makes}, \cite{marlow2013activity}, \cite{trinkenreich2020hidden} & \hfil \faMale \space \faMale \space \faMale \\
\hline

\skillheader{Organization.} Understands the file and team structure of a project & \hfil \cite{hoda2012self}, \cite{kalliamvakou2017makes}, \cite{li2015makes}, \cite{trinkenreich2020hidden} & \hfil \faMale \space \faMale \\
\hline

\skillheader{Processes.} Follows project processes and rules & \hfil \cite{dias2021makes}, \cite{groeneveld2020non}, \cite{hoda2012self}, \cite{kalliamvakou2017makes}, \cite{li2015makes}, \cite{mendez2018open} & \hfil \faMale \space \faMale  \\[8pt]
\hline

\multicolumn{3}{l}{\cellcolor[rgb]{ .921,  .921,  .921}\themeheader{T9}{Interpersonal Skills}}\\
\hline

\skillheader{Kind.} Empathizes with others and show kindness & \hfil \cite{coelho2018we}, \cite{groeneveld2020non}, \cite{kalliamvakou2017makes}, \cite{li2015makes}, \cite{mendez2018open} & \hfil \faMale \space \faMale \space \faMale \\
\hline

\skillheader{Communication.} Clearly communicates with others in written and oral forms & \cite{baltes2018towards}, \cite{coelho2018we}, \cite{groeneveld2020non}, \cite{hoda2012self}, \cite{kalliamvakou2017makes}, \cite{li2015makes}, \cite{mendez2018open}  & \hfil \faMale \space \faMale \space \faMale \space \faMale \space \faMale \\
\hline

\skillheader{Asking for help.} Knows how to ask for help when one needs it & \hfil \cite{groeneveld2020non}, \cite{li2015makes} & \hfil \faMale \\
\hline

\skillheader{Giving help.} Actively helps others when they need it; mentors, guides, or teaches others in the team & \hfil \cite{baltes2018towards}, \cite{dias2021makes}, \cite{groeneveld2020non}, \cite{hoda2012self}, \cite{kalliamvakou2017makes}, \cite{li2015makes} & \hfil \faMale \space \faMale \\
\hline

\skillheader{Conflict resolution.} Resolves conflicts between people &  \hfil \cite{groeneveld2020non}, \cite{kalliamvakou2017makes} & \hfil \faMale \\
\hline

\skillheader{Collaboration.} Works with people from diverse backgrounds to achieve a shared goal & \hfil \cite{groeneveld2020non}, \cite{hoda2012self}, \cite{kalliamvakou2017makes}, \cite{li2015makes} & \hfil \faMale \space \faMale \space \faMale \space \faMale \\[8pt]
\hline

\multicolumn{3}{l}{\cellcolor[rgb]{ .921,  .921, .921} \themeheader{T9}{External Relations}}\\
\hline

\skillheader{Stakeholders.} Works with stakeholders to coordinate OSS development & \hfil \cite{dias2021makes}, \cite{groeneveld2020non}, \cite{hoda2012self}, \cite{kalliamvakou2017makes}, \cite{li2015makes}, \cite{trinkenreich2020hidden} & \hfil \faMale \space \faMale \\
\hline

\skillheader{Marketing.} Promotes the project to broader audiences &  \hfil \cite{hoda2012self}, \cite{trinkenreich2020hidden} & \hfil \faMale \\
\hline

\skillheader{Licenses.} Understands the process of obtaining a license for the OSS project & \hfil \cite{trinkenreich2020hidden} & \hfil \faMale \\
\hline

\skillheader{Funding.} Obtains funding for the OSS project &  \hfil \cite{trinkenreich2020hidden} & \hfil \faMale \\[8pt]
\hline

\multicolumn{3}{l}{\cellcolor[rgb]{ .921,  .921, .921} \themeheader{T9}{Management}}\\
\hline

\skillheader{Community.} Cultivates an engaged, healthy community; is involved in community governance &\cite{coelho2018we}, \cite{dias2021makes}, \cite{groeneveld2020non}, \cite{hoda2012self}, \cite{kalliamvakou2017makes}, \cite{li2015makes}, \cite{trinkenreich2020hidden} & \hfil \faMale \space \faMale \space \faMale \\
\hline

\skillheader{Project.} Is a project leader and directs the completion of a project (e.g., setting standards, meeting deadlines) &  \hfil \cite{dias2021makes}, \cite{kalliamvakou2017makes}, \cite{li2015makes}, \cite{marlow2013activity}, \cite{trinkenreich2020hidden} & \hfil \faMale \space \faMale \space \faMale \\
\hline

\skillheader{Planning.} Plans the next steps for what is ahead & \hfil \cite{dias2021makes}, \cite{li2015makes}, \cite{trinkenreich2020hidden} & \hfil \faMale \space \faMale  \\
\hline

\skillheader{Delegating.} Effectively assigns tasks to others & \hfil \cite{kalliamvakou2017makes} & \hfil \faMale \\
\hline

\skillheader{Time.} Manages one's time effectively given the tasks at hand &\hfil \cite{coelho2018we}, \cite{li2015makes}, \cite{trinkenreich2020hidden} & \hfil \faMale \\[8pt]
\hline

\multicolumn{3}{l}{\cellcolor[rgb]{ .921,  .921,  .921}\themeheader{T9}{Characteristics}}\\
\hline

\skillheader{Adventurous.} Is willing to take risks & \hfil \cite{coelho2018we}, \cite{groeneveld2020non}, \cite{kalliamvakou2017makes} \\ 
\hline

\skillheader{Open-minded.} Is open to new ideas and viewpoints &  \hfil \cite{baltes2018towards}, \cite{dias2021makes}, \cite{groeneveld2020non}, \cite{kalliamvakou2017makes}, \cite{li2015makes} & \hfil \faMale \space \faMale \\
\hline

\skillheader{Patient.} Accepts unexpected delays or problems without being upset & \hfil \cite{baltes2018towards}, \cite{dias2021makes}, \cite{groeneveld2020non}, \cite{mendez2018open} & \hfil \faMale \space \faMale \\
\hline

\skillheader{Adaptable.} Adapts to changes in the environment & \hfil  \cite{groeneveld2020non}, \cite{li2015makes} & \hfil \faMale \\
\hline

\skillheader{Curious.} Is eager to learn or know new things; continuously learns & \hfil \cite{baltes2018towards}, \cite{dias2021makes}, \cite{groeneveld2020non}, \cite{li2015makes} & \hfil \faMale \space \faMale \space \faMale \\
\hline

\skillheader{Reliable.} Is reliable and consistent between one's actions and words & \hfil \cite{groeneveld2020non}, \cite{kalliamvakou2017makes} & \hfil \faMale \\
\hline

\skillheader{Persevering.} Consistently works through difficult or unexpected situations independently and with a positive attitude & \hfil \cite{groeneveld2020non}, \cite{li2015makes} & \hfil \faMale \\
\hline

\skillheader{Diligent.} Works hard to complete tasks; is committed and focused & \hfil \cite{dias2021makes}, \cite{groeneveld2020non}, \cite{kalliamvakou2017makes}, \cite{li2015makes}, \cite{marlow2013activity}, \cite{mendez2018open} & \hfil \faMale \space \faMale\\
\hline

\skillheader{Self-aware.} Assesses one's situation and makes corrective actions; is transparent and accountable to one's actions; displays confidence based on an accurate understanding of abilities & \hfil \cite{baltes2018towards}, \cite{dias2021makes}, \cite{groeneveld2020non}, \cite{li2015makes} & \hfil \faMale \\

\bottomrule

\end{tabular}
}%
 \label{tab:SkillsTable}
\end{table*}

\subsection{Model of Skills in OSS}

\label{sec:SkillsModel}

\paragraph{Technical Skills.} The \category{Technical Skills} category encapsulates technical knowledge and abilities needed to perform a specific task. Many of these skills were technology-related. Basic \subcategory{programming} skills, such as debugging, testing, and code comprehension, were the most commonly mentioned skill in our model. Participants argued that programming skills were central to OSS development: \inlinequo{Being able to write working computer code and tests, kind of the backbone of OSS.}{R_p9FY9CK3d5ODp4d} In this category, participants also frequently mentioned skills in more advanced aspects of \subcategory{software engineering}, including best practices on software design, requirements engineering, and maintenance. Supporting these technical skills was an understanding of how to use \subcategory{version control systems}, which was the backbone for remote collaboration. This knowledge included understanding how to resolve merge conflicts~\cite{gousios2015work} as well as \inlinequo{the flow of clone, add, commit[, and]...rebas[ing] and cherry-pick[ing]}{R_VWhRr8CXRQJtix3}. Furthermore, competency in specific \subcategory{technologies} as well as \subcategory{DevOps} skills promoted efficient software development. Lastly, \subcategory{domain} knowledge, which encapsulated specialized knowledge from fields directly related to and unrelated to technology (e.g., transportation, finance, cryptography, distributed systems), provided valuable expertise which drove OSS development.

\paragraph{Working Styles.} The \category{Working Styles} category describes how an individual does their work. For example, having an \subcategory{organized} working style helped contributors manage repositories, teams, and tasks. One participant noted how this skill contributed to the project being seen as well-maintained: \inlinequo{Skill in organization is important as well; guiding users that file issues with templates that help you get the information you need, labeling and organizing into milestones...can signal to consumers that a project is well maintained.}{R_3n62a0Xyt14glla} 

Participants also expressed an appreciation for contributors who pursued \subcategory{excellence} in their work, where individuals would go above and beyond to produce high-quality contributions. Finally, being \subcategory{available} allowed contributors to give feedback or answer questions to their teammates in a timely manner. Both these skills were frequently cited in literature and were associated with great maintainers~\cite{dias2021makes} and software engineers~\cite{li2015makes}.

\paragraph{Problem Solving.} The \category{Problem Solving} category includes skills related to solving open-ended problems. Contributors described this skill as at the crux of OSS development: \inlinequo{Analytical and problem-solving are more important for building any software.}{R_sGBhdNGbuVb0O0p} For problem solving, being \subcategory{creative} helps contributors to brainstorm innovative solutions to solve the problem, while being \subcategory{analytical} enables contributors to select the best solution through evaluating trade offs of one solution over another. However, a critical component to solving a problem is moving past thinking about the problem and its solutions by having \subcategory{initiative} to execute solving the problem. One contributor summarized the problem solving process succinctly: \inlinequo{[Skills that] enable you to quickly assess the situation, figure out what needs to be done and just do it.}{R_28MS4Ld3m7MWhSf}

\paragraph{Contribution Types.} The \category{Contribution Types} category encapsulates types of contributions to OSS projects. Most of these codes were not explicitly mentioned in the literature sources but were elicited through participants. \subcategory{Bug reporting}, which encapsulates identifying defects in OSS products and providing clear and concise instructions to reproduce it, as well as \subcategory{bug triaging}, which describes parsing through bug reports and prioritize them for the project, were listed as OSS skills. The most common contribution brought up by participants was \subcategory{documentation}, which is essential to teach others how to use OSS or contribute to a project. Another type of contribution was \subcategory{visual design}, which describes competency in designing, prototyping, or developing user interfaces or graphic art. Participants also mentioned contributing feedback through \subcategory{code reviews} to help others improve as a type of OSS contribution. Finally, contributors identified \subcategory{translation} of materials to different languages as a skill, such as \inlinequo{contributing to Brazilian Portuguese translation of repositories.}{R_1E40USJYRn54NO5} The diversity of contributions was vital to OSS, as one participant summarized: \inlinequo{Most programmers suck as \sic documentation, so users and technical writers help fill those gaps. Programmers suck as \sic finding their own bugs, so users are critical to finding and fixing bugs.}{R_2CV87PJRcFAP2T2}

\paragraph{Project-Specific Skills.} The \category{Project-Specific Skills} category is skills related to the particular OSS project. Understanding and following project \subcategory{processes}, such as \inlinequo{submitted pull request has comprehensive description, with requisite automated tests}{R_RIbA9JTcOqm6tvb}. To make useful contributions, participants said that contributors should understand and believe in the project's \subcategory{purpose} and its trajectory: \inlinequo{[One OSS skill is to] create a pull requests \sic that fixes a bug or adds a feature that is in line with the package mission statement and doesn't break any other functionality.}{R_3h6dw6lET34xrMj} This skill also includes whether a contributor understands and is aligned with the broader impact of OSS in the world: \inlinequo{[One OSS skill is to] have the capacity to understand the support you are giving to an emerging community, software has to be something for everyone, everything you give to a project is something that someone needs...that can change the future of the world.}{R_32PXCJs51BmEE4p} Finally, knowing the \subcategory{organization} of a project helps contributors work more effectively by knowing where resources are located in the repository as well as who to ask for help based on the team members' expertise.

\paragraph{Interpersonal Skills.} The \category{Interpersonal Skills} category refers to skills used to interact with others. \subcategory{Communication} enabled \inlinequo{asynchronous and distributed collaboration}{R_3FKEk65eiGg7XjF}, and was one of the most highly cited skills from literature and by participants. \subcategory{Collaboration} was also a common theme from participants. This skill enabled groups of people with \inlinequo{different culture, timezone and skills}{R_2cioupPEFARrY1v} to work with one another. Thus, skills in \subcategory{conflict resolution} helped facilitate teamwork in OSS teams. 

Being \subcategory{kind} was central to interpersonal skills. One participant explained how it facilitated collaboration: \inlinequo{Empathy is also a two-way street; understanding the communication patterns of others and realizing when someone isn't too far gone to be guided back to a path of productive communication is important... It takes skill to work with these people, de-escalate, and help guide them away from toxic behavior and back towards being effective contributors.}{R_3n62a0Xyt14glla} Additionally, participants noted a culture of \subcategory{giving help} and \subcategory{asking for help} as vital to OSS. One participant explained how helping others grows communities: \inlinequo{A bit of mentorship can come a long way in growing the community, if there's more on an OSS project than the product/contents like a knowledge base to tap into.}{R_2B9jRvNxtZolwf0}

\paragraph{External Relations.} The \category{External Relations} category describes skills related towards working with and building relationships individuals outside the OSS team. The most frequently cited skill from literature and respondents under this category was building relationships with \subcategory{stakeholders}. One participant explained how these relationships helped to understand user needs and pain points: \inlinequo{If you are socially skilled, you can contribute to OSS by conducting surveys with the OSS users to highlight pain points and helpful features.}{R_2zC5ofuQV0viNST} Other aspects of external relations included identifying and obtaining proper \subcategory{licenses}, \subcategory{marketing} the project to broader audiences, and finding \subcategory{funding}. The latter two skills were cited as important in promoting the OSS project's sustainability~\cite{dias2021makes}.

\paragraph{Management.} The \category{Management} category refers to skills related to managing OSS projects, teams, communities, or tasks. \subcategory{Planning}, \subcategory{delegating}, and \subcategory{time} management helped contributors execute tasks efficiently and supported other management-related skills. For instance, managing a \subcategory{project} was a popular skill which involved forming a long-term, ``macro'' view of the project and efficiently lead the team and tasks to remain on schedule. Meanwhile, participants mentioned skills in managing the \subcategory{community}, which included being involved in community governance or providing internal support to foster an engaged, welcoming community. One participant explained how this skill was vital to successful OSS communities: \inlinequo{The Rust community is a highly successful FOSS community and...[that] is due to non-technical contributors who...maintain and moderate community spaces.}{R_R8khgvz3arN4oiB} Managing projects and communities were frequently cited in literature and by participants.

\paragraph{Characteristics.} The \category{Characteristics} category includes characteristics or personality traits of OSS contributors. Participants noted some characteristics which facilitated learning and self-improvement. In particular, being \subcategory{open-minded} helped contributors be \inlinequo{open to ideas}{R_1q1AWLuBoMsbmi2}, \inlinequo{open to feedback}{R_29vxy5iVHPopyQ7} and allowed them to \inlinequo{change [their] mind}{R_pDC1DMUXAJJw3UB} when new information was presented. Being \subcategory{curious}, which was the most cited trait in the category, allowed contributors to gain new knowledge and skills as well as \inlinequo{learn from other’s work}{R_SDzTK98b5XqflHH}---one participant even said, \inlinequo{self learning is the key skill in oss.}{R_10wmOY99YAsdGZa}

Character traits which enabled contributors to be pleasant to work with were also important. Participants mentioned that being \subcategory{patient} with themselves and others was important because \inlinequo{OSS maintainers are doing it pro-bono [and] people may be busy with their life}{R_1IK2VRbEXRYw7tF}. Additionally, being \subcategory{adaptable}, \subcategory{self-aware}, and \subcategory{reliable} were listed as examples of OSS skills.

Some characteristics more directly influenced contributors' work. For example, being \subcategory{adventurous} helped contributors to move into high-value areas or experiment with new ideas. Being \subcategory{persevering} enabled contributors to \inlinequo{figure things out to the best of [their] ability without asking others for help immediately}{R_3CBOk54KYzln35s}, while being \subcategory{diligent} improved one's commitment to and passion for the project.

\subsection{Validation of the Model}

To validate the skill model, we collected a large list of popular skills related to software development and checked if the skills can be mapped to the skill model.
To collect the popular skills, we used LinkedIn, an online platform that allows job seekers to post their CVs and employers to post jobs. With the LinkedIn Talent Insights tool, we collected the 100 most common skills for three talent pools:

\begin{itemize}
    \item \emph{Software engineers}. Professionals with a software engineering job title (e.g., `Software Engineer'', ``Senior Software Engineer'', ``Lead Software Engineer''). %
    \item \emph{General software development}. Professionals who listed ``Software Development'' as a skill in their LinkedIn profile.
    \item \emph{Open source}. Professionals who listed ``Open Source Software'' or ``Open Source Development'' as skills.
\end{itemize}

We combined the three lists and removed duplicates, resulting in a dataset of 157 popular skills that professionals use on LinkedIn.

Next, to \emph{validate} the skill model, we mapped every LinkedIn skill to codes in our skill model. We were able to map all LinkedIn skills to our model without having to create new categories. This suggests that the skill model is adequate to capture popular software development skills on LinkedIn. 

Most of the LinkedIn skills (49.7\%) were mapped to \subcategory{Technologies} in  \category{Technical Skills} and no LinkedIn skills were mapped to \category{Characteristics},  \category{Interpersonal Skills}, \category{Working Styles}, and  \category{Project-Specific Skills}. This is \textbf{not} a limitation of our skill model; it rather shows that LinkedIn is a site where professionals seem to share technical skills more than other skills. This highlights opportunities for showcasing skills from the other categories as we will discuss in Section~\ref{subsec:collaboration}.

\mybox{\faArrowCircleRight~\textbf{Summary:} Skills in OSS comprise of a mix of hard and soft skills. Major categories of skills are  technical skills, working styles, problem solving, contribution types, project-specific skills, interpersonal skills, external relations, management, and characteristics.}

\begin{table}[]
    \centering\small

 \caption{Reasons why participants learned new skills with over 20 occurrences; for the full list, please refer to the supplemental materials~\cite{supplemental-materials}.}
 \vspace{-0.5\baselineskip}
    \begin{tabular}{p{0.82\linewidth}p{0.10\linewidth}}
        \toprule
        \textbf{Reason} & \textbf{Count} \\
        \midrule
        Amount of resources available or difficulty to learn the skill & \hfil 68 \\
        Benefitting themselves or professional career in the future & \hfil 56 \\
        \hangindent=1em Importance and usefulness of the skill on the OSS community, teammates, and world & \hfil 45 \\
        \hangindent=1em Improving areas of weaknesses identified by the contributor or others & \hfil 42 \\
        Personal interest in an area or project that requires the skill & \hfil 39 \\
        Personal goals & \hfil 28 \\
        What is required for their current roles in OSS projects & \hfil 28 \\
        \bottomrule
    \end{tabular}

    \label{tab:why-learn-skills}
\end{table}

\section{Improving skills (RQ2)}
\label{sec:improving}

In this section, we present the findings for the research question: \textbf{``\RQB''}

\paragraph{Why contributors improve skills.} In our survey, we asked participants to explain what factors influence improving skills. We enumerate a subset of these factors in Table~\ref{tab:why-learn-skills}. Participants most commonly said they learned new skills based on their interests; what would benefit their professional career; the needs of the OSS community, teammates, or the world; resources available to them, such as time and energy; and what they or others identified as potential areas of improvement.

Participants also explained how the impact of the skill was a factor: \inlinequo{I primarily do open source because it makes me feel good and I love solving problems. I don't think about my OSS work in terms of skills. I think in terms of how can I make the world a better place.}{R_2CV87PJRcFAP2T2} Some participants said they would do a cost-benefit analysis of learning the skill based on the resources required to learn the skill and its impact: \inlinequo{Time availability vs impact - I have almost no time, so everything I do has to have a point, sadly.}{R_41MLZXd8lD21XoZ} Other factors for learning skills included the contributor's personal goals, their interest in learning for self-improvement,  their current role, personal circumstances, trends in the field, their working environment (e.g., having an inclusive community), and their previous experience. Some participants reported not having a rationale to learn skills. Additionally, some said learning skills expanded their social network: \inlinequo{Growing social contacts. I already have 16000 followers on LinkedIn and would like to grow it more.}{R_1q1AWLuBoMsbmi2}

A few contributors mentioned they were experienced and had little need to improve their skills. \inlinequo{My number of years of OSS support (over 25), as well as my advanced age (77) are such that I probably won't be changing my ways very much improvement-wise.}{R_1fd7wE31tWZI8tl}

\begin{table}[]
    \centering\small

 \caption{Examples of how participants considered growing their skills with at least 40 occurrences; for the full list, please refer to the supplemental materials~\cite{supplemental-materials}.}
 \label{tab:how-grow-skills}

 \vspace{-0.5\baselineskip}
    \begin{tabular}{p{0.82\linewidth}p{0.10\linewidth}}
        \toprule
        \textbf{Example} & \textbf{Count} \\ %
        \midrule 
        Continued efforts to keep learning & \hfil 90 \\
        Contributing \& helping maintain a project & \hfil 86 \\
        Actively practicing skills to grow & \hfil 83 \\
        Reading code committed by others & \hfil 54 \\
        Reading non-code content including blogs and documentation & \hfil 64 \\
        Reading over code reviews as well as reporting \& fixing bugs  & \hfil 63 \\
        Actively engaging with the project and community & \hfil 61 \\
        Consistently challenging oneself & \hfil 47 \\
        Observing experts and finding mentors across projects & \hfil 44 \\
        Networking \& forming strong connections w/ peers & \hfil 44 \\
        Staying curious \& creative by tinkering  & \hfil 43 \\
        \bottomrule
    \end{tabular}

\end{table}

\paragraph{How contributors improve skills.}
Participants described a host of strategies used to grow their OSS skills, ranging from independently working to collaborative tasks. We list a subset in Table~\ref{tab:how-grow-skills}.
For instance, participants mentioned learning by reading blogs, following public OSS figures on social media, and staying up-to-date with conferences. 
Another source participants referenced for being updated on the latest frameworks was checking \inlinequo{open source hotspots}{R_tDnYbecmCAS03gB}, such as \gh's trending topics page. In these settings, participants described how helpful it was to learn from other people: \inlinequo{Learn from other people. People that work in such environments may be a wonderful example of how to do that kind of things. Watch how other people work in order to design the workflow that fits you the best.}{R_3lVOwui1uqPUXda} In addition to following others on external platforms, participants also found it extremely valuable to read code, close issues, and merge pull requests to grow their skills.

Some participants also described taking their approaches to improve a step further by tinkering and engaging with the community where they can: \inlinequo{Contribute to open source projects, there are countless projects on \gh in most languages/areas where there really is something to get involved in from beginner to advance \sic.}{R_25t1KSlmYEvXaOF}
Participants also described that a great way to improve OSS skills is to be consistent about practice by \inlinequo{contribut[ing] every day}{R_29nXl92y6rCPciO} and challenging themselves to step outside their comfort zone and collaborate with new people on new projects: \inlinequo{Gain experience in different technical projects to up my technical expertise. Work with different people, possibly total strangers remotely to practise good communication skills.}{R_1QskZirgKCsUo7U}
By the same token, participants described that nothing beat getting their hands dirty with making contributions, no matter their level of expertise: \inlinequo{Start small, even if you are an expert. Take some easy and maybe a little bit boring issues at first so you can grasp the vision of the project. Then, start to work progressively on harder tasks, ask for feedback and discuss. %
}{R_2tMty22IqpCKjnC}

\paragraph{Contributor thoughts on improving skills.} When asked questions around participants' future plans to improve skills, the majority of participants knew how to improve them (72\%) and were taking steps to improve (72\%) the OSS-related skills they want to grow. 

We asked participants if they joined projects based on the skills that they could learn (see Table~\ref{tab:join-projects}). Learning new skills is important: 50\% chose projects based on skills that they can gain and 42\% chose projects based on project members who are strong at a particular skill. Giving back to projects is important too, if not more: 80\% chose projects based on the skills that they can offer and 69\% chose projects that are looking for contributors with the skills they have.

\mybox{\faArrowCircleRight~\textbf{Summary:} Contributors improve skills by reading blogs, following OSS figures on social media and conferences, learning from peers, and contributing to OSS projects. Growing skills is influenced by contributor interests, needs of the OSS team, resources available, and identified areas of improvement. Most contributors are actively taking steps to improve their skills and join projects based on skills they have to offer.}

\begin{table}[]
    \centering%
 \caption{Percentage of respondents who contribute to an OSS project based on what they can offer vs.~ what they gain.}
 \vspace{-0,5\baselineskip}
\resizebox{\linewidth}{!}{%
    \begin{tabular}{lc}
        \toprule
        I choose to contribute to an OSS project based on\dots & \\
        \hspace{4mm}the skills that I can offer to the project.	 & 80\% \\
        \hspace{4mm}the skills that I can gain from the project. & 50\% \\
        \midrule
        I would choose to contribute to a project if I knew\dots 
        \\
        \hspace{4mm}they were looking for contributors with a skill I have.  & 69\% \\
        \hspace{4mm}it had a collaborator who was strong at a particular skill. & 42\% \\
\bottomrule
    \end{tabular}
} %
   
    \label{tab:join-projects}
\end{table}

\begin{table}[]
    \centering\small

 \caption{Motivations for why participants would share their skills with others with over 20 occurrences; for the full list, please refer to the supplemental materials~\cite{supplemental-materials}.}
 \vspace{-0.5\baselineskip}
    \begin{tabular}{p{0.82\linewidth}p{0.10\linewidth}}
        \toprule
        \textbf{Motivation} & \textbf{Count} \\
        \midrule
        Improving career prospects & \hfil 58 \\
        Building a profile of who they are as a contributor  & \hfil 55 \\
        Promoting transparency and building trust with others & \hfil 49 \\
        Improving reputation and self-confidence through recognition & \hfil 43 \\
        Finding experts or being considered as an expert by others & \hfil 42 \\
        Finding new opportunities to contribute to on OSS platforms & \hfil 39 \\
        Becoming motivated to improve themselves or others & \hfil 30 \\
        Helping others, especially via knowledge sharing & \hfil 27 \\
        Expanding social network & \hfil 25 \\
        Comparing to others and ``gamifying" the skills system & \hfil 23 \\
        \bottomrule
    \end{tabular}

    \label{tab:why-share-skills}
\end{table}

\section{Showing skills (RQ3)}
\label{sec:showing}

In this section, we present the findings for the research question: \textbf{``\RQC''}

\paragraph{Why contributors share skills.} We asked participants what motivates them to share skills with others. A subset of reasons are in Table~\ref{tab:why-share-skills}. Participants reported wanting to share their skills mainly for themselves: to improve career prospects, showcase expertise, build a profile of their contributions, improve their self-confidence and reputation, and receive recognition. Another major factor for sharing skills was for transparency to build trust in the OSS community. Some contributors mentioned how transparency allowed them to see how others were growing new skills: \inlinequo{It's good to see what do different people like, what they're focusing on, how they're trying to grow, etc.}{R_29jqIQhEE6FJPiv} Contributors also mentioned other motivations, such as finding experts in specific skills, facilitating collaboration, helping others by sharing knowledge, being matched to projects, building communities around a skill, motivating themselves and others to improve, and competing against other contributors.

Notably, some participants also mentioned how they wanted to have their skills be shown as proof they were qualified in their professional career, indicating that skills, if automatically detected, should be accurate: \inlinequo{[I would show my skills] if they are objective and would provide some more weight to my actions.}{R_24qv3KbVmXGLTUE}

Additionally, in the survey, 51\% of participants said that they would be more motivated to improve their OSS-related skills if they were displayed publicly on \gh.

\paragraph{How contributors show their skills.} We asked participants if and how they currently display skills relevant for OSS development. \gh was used by 74\% of participants and LinkedIn was used by 54\%. On \gh, participants displayed their skills through their public repositories (90\%), through active contributions to the OSS projects (81\%), and through the \gh profile description (62\%). On LinkedIn, participants displayed skills in the Skills section in the profile (80\%), by including projects in their profile (61\%), and in the Experience section (56\%). 
For both \gh and LinkedIn, several participants mentioned including links to other online profiles.

\begin{table}[]
    \centering\small

 \caption{Examples of when participants consider it to be useful to share their skills with at least 20 occurrences; for the full list, please refer to the supplemental materials~\cite{supplemental-materials}.}
 \label{tab:when-share-skills}

 \vspace{-0.5\baselineskip}
    \begin{tabular}{p{0.82\linewidth}p{0.10\linewidth}}
        \toprule
        \textbf{Example} & \textbf{Count} \\ %
        \midrule 
        Recruiting and finding jobs & \hfil 107 \\
        Assessing projects, contributors, and pull requests & \hfil 49 \\
        More effective collaboration & \hfil 44 \\
        Participating or joining a community. Being a new contributor. & \hfil 38 \\
        Finding experts and contributors for a task or project & \hfil 34 \\
        Establish credibility & \hfil 27 \\
        \bottomrule
    \end{tabular}

\end{table}

\paragraph{When contributors share skills.} We also asked participants about examples of when it could be useful to share skills with others. The most prominent examples were related to recruiting and personal branding, i.e., showing their portfolio of work online to help recruiters discover contributors: \inlinequo{Displaying skills next to projects and contributions would be a great all-in-one CV for developers, designers and other tech people}{R_2tMty22IqpCKjnC} Recruitment was mentioned both in the context of industry work but also in the context of open source, e.g., current project members could use skill information \inlinequo{to decide [to] include or not a person as a collaborator}{R_2zT3HOKdHV4EuVU}.

Other examples that were mentioned related to assessing one's contributions to an open source project. Skill information can be useful for pull requests, both for the reviewer to understand the submitter's background and for the submitter to get their voice heard:
\inlinequo{To offer some legitimacy / street cred when seeking to get a PR merged into a project}{R_3dS8kc3qA5yIYTj}
and
\inlinequo{I think it's helpful in having people listen. The way having a lot of commits to certain projects can make people care about what you have to say.}{R_2dsEJCtPlNFq37S}

Skill information was also considered useful for newcomers. Knowing someone's skills can help existing members to better understand and support new members through mentorship.
\inlinequo{If a certain person is contributing to an Android project, but they have a badge which shows they're newcomers to the Android platform, we could give them more detailed, step-by-step feedback or explain some concepts using simpler terms when reviewing}{R_29jqIQhEE6FJPiv}
Another participant explained:
\inlinequo{For example, if I’d want to grow my skills on the Elixir language, and someone I know notices it, they might offer me a pair programming or sparring session on one of their projects. I might also do the same for people around me.}{R_XGL2exJfm8mKPbb}

Other examples were centered around collaboration between collaborators, specifically for expert finding: \inlinequo{Facilitate the search for people with the right skills.}{R_2zcrg66yY5pW8MO}
Learning was mentioned as another example. Participants suggested that skill information can help people to find other contributors to learn from:
\inlinequo{If a contributor wants to learn how to contribute to open source, looking at the profile of someone who's made lots of contributions would be helpful}{R_pGJc9kMnkEOPGA9}

\paragraph{What skills information contributors would share.} Table~\ref{tab:display-information} shows the willingness of participants to share skill-related information. The majority of participants would share their skills (76\%) and activities related to the skills (60\%) with the public. For the skill ratings, the majority would share with their collaborators (63\%). Most participants would keep ways to improve skills private and not share with anyone (58\%). It is noteworthy that 30\% of participants would be willing to share all four information types with the public. As one participant put it: \inlinequo{If [skills] are based on real, objective data, I don't see why hiding them. Unless that involves classified work or things that can get you in prison or worse troubles.}{R_3snR3zX6LOBxkpl}

\paragraph{How maintainers assess contributor skills.} Amongst maintainers, 59\% of participants used \gh to evaluate what skills a potential contributor has. The main information source was public \gh activity (97\%). They also looked at public repositories owned by the contributor (83\%), contributions of the contributor on the maintainer's projects (79\%) and other projects (64\%), interactions with the contributor on the maintainer's projects (60\%) and other projects (44\%), personal websites (54\%), and social media activity (34\%).

\paragraph{Feedback on sharing skills.} Some participants expressed strong concern for sharing skills on OSS platforms, citing that it could create a gamified or competitive environment that would not show a holistic view of contributors' abilities. One participant warned that it would make \inlinequo{\gh a replacement for LinkedIn.}{R_tJrQIk6s1eEnZC1} Another expressed distaste towards sharing skills if skills were automatically detected: \inlinequo{I absolutely hate the idea that people's behaviour and priorities are going to be influenced by artificial gamification metrics, rather than trying to be doing the best they can with reasonable flexibility for the projects we should all care about, and helpful and kind to maintainers.}{R_31tvWn9Ve6tws4K} However, others felt that automatic detection of skills could be beneficial: \inlinequo{It's a small psychological reward for the effort I put in.}{R_Ce6WM4ItvLS4wud} and \inlinequo{I visit \gh multiple times a day, so \gh should just infer [skills] from work and show callouts every week/something for sharing confirmation.}{R_1eFv96abv8vhPnZ}

\medskip
\mybox{\faArrowCircleRight~\textbf{Summary:} Contributors show their skills and maintainers assess potential contributors on \gh. Contributors share skills to improve career prospects and showcase expertise. Contributors find skills useful for recruitment, assessing contributions, orienting newcomers to the community, and finding experts.}

\begin{table}[]
    \centering%

 \caption{Willingness to share skill-related information on GitHub with the public and with collaborators. Bold values indicates the majority response.}
 \vspace{-0.5\baselineskip}
    \begin{tabular}{@{\phantom{x}}lccc@{\phantom{x}}}
       
        \toprule
        & \multicolumn{2}{c}{Share with} & \\
        \cmidrule{2-3}
        Type of Information & Public & Collaborators & \stackanchor{Keep}{Private} \\
        \midrule
        Skill name & 76\% & \bf 85\% & 15\% \\
        \includegraphics[scale=0.32]{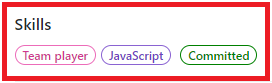} \\
        \midrule
        Activities related to the skill & 60\% & \bf 82\% & 18\% \\
        \includegraphics[scale=0.32]{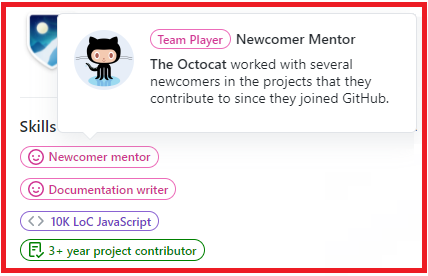} \\
        \midrule
        Skill rating & 47\% & \bf 63\% & 37\% \\
        \includegraphics[scale=0.32]{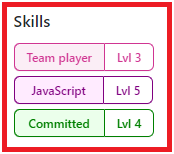} \includegraphics[scale=0.32]{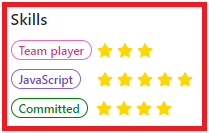} \\
        \midrule
        Ways to improve the skill & 30\% & 42\% & \bf 58\% \\
        \includegraphics[scale=0.32]{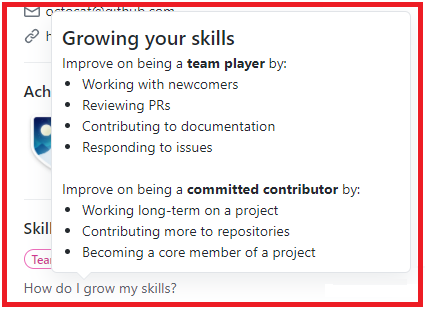} \\
        \bottomrule
    \end{tabular}

    \label{tab:display-information}
\end{table}

\section{Discussion}
\label{sec:Discussion}

Our findings indicate that skills should be a first-class citizen in software engineering and OSS development. Skills could help improve many existing tools and experiences (e.g., pull requests) or create new experiences (e.g., mentorship). However, this requires careful thought on how to incorporate skills in platforms and tools. 
Skills could also enrich empirical studies, for example, for project success and team composition.
Below, we synthesize our findings with prior literature and discuss implications of this work.

\subsection{Detecting Skills}
\label{sec:DetectingSkills}
Our skills model (see Section~\ref{sec:SkillsModel}) indicates that both soft and hard skills are crucial to OSS, which is corroborated from previous work~\cite{vadlamani2020studying}. Prior work has automatically detected expertise and skills from online sources, such as \gh~\cite{montandon2021mining, hauff2015matching, greene2016cvexplorer, papoutsoglou2019extracting, montandon2019identifying, siegmund2014measuring,mockus2002icse,Zhou:2010,bergersen2014construction}, but largely focuses on software engineering-related skills.

An open problem remains: how do we automatically mine soft and hard skills from online data? 
Answering this question is a prime research direction for mining software repositories. Prior work has investigated this issue~\cite{liang2022towards}, but more research is needed.
From our qualitative analysis, some participants viewed detected skills as a source of truth to assign credibility to their actions. Since \gh users use profiles to judge individuals and their contributions~\cite{marlow2013impression, ford2019beyond}, this could impact \gh activity, such as whose contributions are accepted.
Thus, methods to detect skills should be accurate, objective, and grounded in developer activity traces.
Research in this area should carefully consider holistic measures of skill detection to be inclusive of newcomers.
For example, being \subcategory{diligent} may be shown through continuous contributions made over a long duration, a high volume of contributions, or responsiveness to assigned issues.

\subsection{Skills Development}
\label{sec:SkillsDevelopment}

Our study indicates that contributors are motivated to learn because a majority of participants are trying to improve their skills in the next 12 months. In software engineering, many individuals use massive open online courses (MOOCs) to learn new skills. MOOCs attract learners who do not aspire to earn formal degrees, but rather gain knowledge that will help them in their roles~\cite{dasarathy2014past}.

OSS platforms may help support this learning since study participants indicated that they learned skills to improve them for OSS and professional settings. This provides an opportunity to extend learning to include soft skills, as opposed to only technical knowledge. This could be increasingly important with the introduction of neural models that generate code with high quality, such as \gh Copilot~\cite{github2022copilot}. Future work could detect skills being learned and recommend supplemental resources to help contributors to learn those skills. Because participants reported continuous practice being crucial to learning skills, future tools could also create individualized curriculum which set ``bite-sized" goals every day to encourage contributors to learn new skills. Additionally, revealing global trends of skills across many OSS projects may help contributors identify skills to improve on based on their potential utility.

\paragraph{Project Matching} Our findings imply that skills are important for matching contributor to projects. OSS contributors are motivated to learn skills needed by projects and are likely to contribute to a new OSS project based on whether the project needs their skill set. Prior work supports this as well---contributors whose motivation was associated with career advancement was correlated with higher levels of satisfaction in participating in a project~\cite{wu2007empirical}. Additionally, common barriers for OSS newcomers are a lack of hard skills, such as software engineering background and domain expertise, as well as soft skills, such as communication, patience, and proactivity~\cite{steinmacher2015social}. Skills could be useful data to match contributors with specific skill sets to projects that require them, potentially reducing barriers to newcomers' first contributions and increasing satisfaction.

\subsection{\gh Profiles}
\label{subsec:profiles}

Incorporating skills into \gh profiles could provide an additional signal to assess OSS contributors and enable profiles to include more dynamic sources of information. Previous work has shown that OSS contributors analyze \gh profiles to assess contributors and their potential contributions~\cite{ford2019beyond, marlow2013impression}. In our study, some participants reported largely being motivated to share skills to market themselves and their expertise, while others wanted additional information to evaluate potential contributors or find experts with specific skills. One could consider revealing some information on skills that contributors are currently learning or improving on for transparency, to be inclusive of newcomers, and showing a contributor's trajectory. Allowing contributors to display skill dimensions that they want to grow (e.g., in a spider graph on their \gh profile) is a possible strategy to indicate what motivates contributors (e.g., a desire to improve their working styles).

Another potential application of skills in \gh profiles could be \gh's achievement badges~\cite{github-badges}, which currently only recognizes contributions to specific repositories or participation in \gh programs. There is an opportunity for OSS platforms such as \gh to extend achievement badges to include OSS skills, using the aforementioned mining techniques (see Section~\ref{sec:DetectingSkills}) to surface them. Similarly, \gh could develop a certification of skills that could be made directly available in a user's profile.

However, our findings also indicate that incorporating skills into \gh profiles requires careful design considerations. Techniques to gamify skills, such as including skill level, could create unhealthy competition and add additional barriers to entry for newcomers. One way to address this concern is by only using gamification for a contributor's self-growth (see Section~\ref{sec:SkillsDevelopment}). Also, consider the access of skills-related information---based on our results, keep the skill name, activities related to the skill, and the rating accessible to only collaborators and ways to improve the skill private. Finally, giving users control over what skills are seen, such as an ability to add or remove skills, may also help reduce this concern.

\subsection{Collaboration}
\label{subsec:collaboration}

Peer feedback is delivered through code reviews in OSS. Code reviews provide technical feedback and assist contributors in peer impression formation, such as understanding each others' expertise and quality of contributions~\cite{bosu2013impact}, as well as knowledge sharing~\cite{bosu2016process}. However, tools that support an individual's OSS skills growth do not exist. These could supplement the feedback provided during code reviews to facilitate expertise-finding and promote contributor growth and high-quality contributions. For example, future work could design tools for OSS contributors to provide feedback on their collaborators' skills, including strengths and areas of improvement. Tools that provide feedback on skills for teams exist, such as OnLoop \cite{onloop2022}, but do not tailor their experiences for OSS contexts. Such tools could be useful for maintainers or mentors to help their team grow their skills. Our work lays the foundation for such a tool by providing a vocabulary to discuss OSS skills.

\paragraph{Mentorship}
Many challenges exist within OSS mentorship. Newcomers struggle with finding mentors~\cite{steinmacher2015social} and defining professional goals~\cite{balali2018newcomers}. Mentors may face language barriers, lack interpersonal skills, or struggle to deliver constructive feedback based on the mentee's background~\cite{steinmacher2021being, balali2018newcomers}. Using skills to match mentors and mentees could increase alignment of mentors' skill sets with mentees. Surfacing mentees' skills could also help mentors craft appropriate feedback, while enabling mentees to identify areas of improvement and set personal goals. Following the design of \gh Discussions Insights Dashboard~\cite{githubviewing2022}, which features newcomers' and daily contributors' discussion activity, as well as a movement to recognize all OSS contributions~\cite{allcontributors2022}, such tools could highlight mentees' valuable non-code contributions.%

\begin{figure}
    \begin{tcolorbox}[left=-14pt,right=2pt,top=2pt,bottom=2pt]
    {\sc \hspace{16pt}Best Practices}
    \begin{itemize}
        \item[{\tiny\faCheck}] Thoroughly validate that the methods to detect skills are accurate and are grounded in contributor activity.
        \item[{\tiny\faCheck}] Design holistic measures of skills, such as thinking of the many ways that a skill may manifest.
        \item [{\tiny\faCheck}] Consider showing what skills a contributor is working on.
        \item[{\tiny\faCheck}] Be careful how skills are gamified.
        \item[{\tiny\faCheck}] Give people control over which skills are shown publicly and to collaborators.
        \item[{\tiny\faCheck}] Reveal what skills are useful or needed in the OSS community.
    \end{itemize}
    \end{tcolorbox}
    \vspace{-0.5\baselineskip}
    \caption{Best practices for skills in OSS.} %
    \label{fig:bestpractices}
\end{figure}

\section{Threats to Validity}
Below, we discuss the threats to validity of our study.

\paragraph{Internal validity}
Survey participants may misinterpret wording or not respond in a way the question intends. To reduce this threat, we had an external researcher review the survey to clarify wording.

Additionally, the structure of the survey may have caused unintentional priming in study participants, especially for what skills in OSS may look like. To address this, we put the question that was used to validate the skills in OSS model in the beginning of the survey before any examples of skills were introduced.
\vspace{-.5em}
\paragraph{External validity}
How generalizable single-case empirical studies, such as this study, to other populations is unknown. However, these studies contribute to scientific advancement~\cite{flyvbjerg2006five}. Our survey includes participants from many OSS roles, genders, ethnicities, and geographic regions, which may improve the generalizability of our findings. Additionally, our sampling approach was designed to capture a wide range of contributions, as it included stratified samples across different OSS project types (OSS, OSS4SG) and contribution types (commits, pull requests, issues, and comments).

Surveys by nature also often suffer from selection bias, and ours is no different. In this study, non-response bias may have occurred because our survey was written in English, which may cause less representation in regions where English is not the primary language. Self-selection bias may also have occurred. Because the survey was marketed as a ``Survey on Skills in OSS'', participants that felt more confident about their skills and contributions in OSS may be more likely to participate in our study. To reduce the effects of the selection bias, we made the survey as short as possible, were transparent about the survey's length, provided incentives to participate in our study, and kept survey responses anonymous.
\vspace{-.5em}
\paragraph{Construct validity}
Evaluating our model using only LinkedIn skills may not be completely representative of skills that practitioners find important in practice. To address this, we will run an additional evaluation of our OSS skills model with OSS experts.

\section{Conclusion}
Skills remain fundamental to the success of OSS development. The diversity of roles and contributions in OSS requires equally diverse skill sets which span both soft and hard skills. In this paper, we investigated the skills that support the development of OSS. We contributed a model for skills in OSS with 45 skills in 9 categories, survey results on how contributors grow their skills and how they would like their skills to be presented, and design implications and best practices on incorporating skills into OSS tools and platforms. %

Our results indicate many future directions for skills in OSS. Contributors may use the results from this study to drive the development of their own skills, while researchers may extend this work to build new tools and experiences or promote further study of skills in OSS.
To facilitate replication of this work, the survey instrument and codebook are available as supplemental material~\cite{supplemental-materials}.

\begin{acks}
We thank our survey participants for their incredible insight on their OSS skills. We also thank Christian Bird, Mala Kumar, Victor Grau Serrat, and Lucy Harris for their feedback. Jenny T. Liang conducted this work for an internship at Microsoft Research’s Software Analysis and Intelligence in Engineering Systems Group. %
\end{acks}

\bibliographystyle{ACM-Reference-Format}
\bibliography{sample-base}

\end{document}